\documentclass[12pt]{article}

\usepackage{amsmath,amssymb,amsbsy,amsfonts,latexsym,graphicx}
\usepackage{color,array,cite}

\numberwithin{equation}{section}

\allowdisplaybreaks

\evensidemargin \oddsidemargin

\begin{document}


\begin{titlepage}

\renewcommand{\thefootnote}{\fnsymbol{footnote}}


\begin{flushright}
\end{flushright}

\vspace{15mm}
\baselineskip 9mm
\begin{center}
  {\Large \bf Complete type IIA superstring action \\
  on IIA plane wave background}
\end{center}

\baselineskip 6mm
\vspace{10mm}
\begin{center}

 Jaemo Park$^a$\footnote{\tt jaemo@postech.ac.kr} and
 Hyeonjoon Shin$^b$\footnote{\tt hyeonjoon@postech.ac.kr}
\\[5mm]
 {\sl $^a$Department of Physics \& Center for Theoretical Physics (PCTP),\\
POSTECH, Pohang 790-784, South Korea }
\\[3mm]
{\sl $^b$Department of Physics, Pohang University of Science
  and Technology (POSTECH),\\
  and Asia Pacific Center for Theoretical Physics (APCTP), \\
  Pohang  790-784, South Korea }

\end{center}

\thispagestyle{empty}

\vfill
\begin{center}
{\bf Abstract}
\end{center}
\noindent We construct the type IIA Green-Schwarz superstring action
on a ten-dimensional IIA plane wave background with 24
supersymmetries keeping the full 32 fermionic coordinates. Starting
from the symmetry superalgebra for the maximally supersymmetric
eleven dimensional plane wave background, we obtain the eleven
dimensional superfields.  The Kaluza-Klein reduction leads to the
ten dimensional superfields for the IIA plane wave background, from
which the type IIA superstring action is constructed.  We show that
the superstring action reduces correctly to the previously known
light-cone gauge fixed action upon imposing the light-cone
$\kappa$-symmetry fixing condition.
\\ [5mm] Keywords : superstring action, plane wave, Kaluza-Klein reduction

\vspace{5mm}
\end{titlepage}

\baselineskip 6.6mm
\renewcommand{\thefootnote}{\arabic{footnote}}
\setcounter{footnote}{0}

\section{Introduction}

Recent years have seen the tremendous development in our
understanding of AdS$_4$/CFT$_3$ correspondence. Important inflection
point is proposal by J. Schwarz \cite{sch} that underlying CFT$_3$
can be written as Chern-Simons matter (CSM) theories without the
usual kinetic term for the gauge fields, especially for higher
supersymmetric theories with ${\mathcal N}\ge4$.
Once that proposal is realized
in specific examples, the understanding of AdS$_4$/CFT$_3$
correspondence has grown by leaps and bounds. After the realization
that the Bagger-Lambert-Gustavsson theory \cite{BL1, BL2, BL3, gus1,
gus2} can be written as the usual $SU(2)\times SU(2)$ Chern-Simons
matter theory \cite{rams}, there appeared a paper by Gaiotto and
Witten \cite{GaiottoWitten} where the attempt is made to write down
${\mathcal N}=4$ Chern-Simons matter theory with matter hypermultiplets.
The attempt was generalized in \cite{HLLLP} which includes twisted
hypermultiplets as well, thereby writing down the general classes of
${\mathcal N}=4$ Chern-Simons matter theories. The special case of such
construction is the famous ${\mathcal N}=6$ theory, known as ABJM theory
\cite{ABJM, HLLLP2}, describing coincident M2 branes on
$C^4/{\mathbf Z}_k$
where $(k,-k)$ is the Chern-Simons level for two gauge groups of
ABJM theory. If the level $k$ is taken to be infinite, the
correspondence becomes that between the ABJM theory and type IIA
superstring theory on AdS$_4 \times CP^3$.

For the further study of the AdS$_4$/CFT$_3$ correspondence, the first
step in the bulk side is to have the IIA superstring action
on AdS$_4 \times CP^3$, which has been constructed based on the super
coset formulation in \cite{Arutyunov:2008if}.  However, as pointed out
already in the coset formulation and emphasized in a subsequent
work \cite{Gomis:2008jt}, the constructed superstring action contains
only 24 fermionic components although type IIA superstring has 32
fermionic components, and thus is not complete.  This problem has been
cured by incorporating the missing 8 fermionic components through
a suitable extension of the coset superspace, and the complete
type IIA superstring action on AdS$_4 \times CP^3$ has been constructed
in \cite{Gomis:2008jt}.

The complete type IIA superstring action opens up the way of
investigating various possible string configurations.  However, it
has rather complicated structure, which may make the full
understanding of string or the quantization of string on the AdS$_4
\times CP^3$ background not so easy. In this case, some simple
background would be helpful. One typical and successful example of
such simplification may be the Penrose limit of the AdS$_5 \times
S^5$ background which leads to the IIB plane wave background
\cite{Blau:2001ne,Blau:2002dy}. The type IIB superstring action
constructed on this IIB plane wave background has turned out to be
exactly solvable \cite{Metsaev:2001bj}, and triggered the
significant progress in the study of AdS$_5$/CFT$_4$ correspondence
\cite{Berenstein:2002jq}.  One may expect that a similar simplified
theory can be constructed also in the present case.  Indeed, it has
been shown in \cite{Nishioka:2008gz,Gaiotto:2008cg,Grignani:2008is}
that a suitable Penrose limit of the AdS$_4 \times CP^3$ background
leads to the previously known IIA plane wave background
\cite{Bena:2002kq,Sugiyama:2002tf,Hyun:2002wu}.  Similar to the IIB
case, the type IIA superstring theory on this background is also
solvable and can be quantized \cite{Hyun:2002wp}.  This manageable
situation through the simplification has been exploited in various
works on the study of AdS$_4$/CFT$_3$ correspondence
\cite{Astolfi:2008ji,Astolfi:2009qh,Agarwal:2010tx,AliAkbari:2010rs,Astolfi:2011ju,Astolfi:2011bg}.
We expect this type of investigation would lead to better
understanding of the stringy aspects of AdS$_4$/CFT$_3$
correspondence.

 In this paper, we revisit the type IIA Green-Schwarz
superstring action on the IIA plane wave background and try to
construct the complete action containing all the 32 fermionic
components. Basically, the motivation comes from the fact that the
IIA plane wave background is not maximally supersymmetric. As
pointed out in \cite{Gomis:2008jt} and explored in more detail in
\cite{Grassi:2009yj}, when the superstring action on a less
supersymmetric background is considered, the $\kappa$-symmetry
fixing condition should be chosen carefully in studying a particular
motion of string.  For example, the type IIA superstring action on
AdS$_4 \times CP^3$ constructed based on the super coset does not
give the correct description of string moving only in AdS$_4$ space,
because it fixes from the beginning 8 fermionic components
corresponding to the broken supersymmetries though those components
play important role in describing such string motion.  In this
sense, the complete action containing all the 32 fermionic
components is required for studying various possible string
configurations or motions.

In the previous construction of type IIA superstring action on IIA
plane wave background \cite{Sugiyama:2002tf,Hyun:2002wu}, the
Penrose limit was taken for the membrane action on AdS$_4 \times
S^7$ \cite{deWit:1998yu}, and the double dimensional reduction was
performed after taking the light-cone gauge fixing condition in
eleven dimensions for simplicity.  We could take the same steps for
obtaining the complete action without any gauge fixing.  In this
paper, however, we will take a different route. Starting from the
symmetry superalgebra of eleven dimensional plane wave background,
the eleven dimensional superfields will be obtained by following the
recipe conceived in \cite{Kallosh:1998zx}.  Then, through the
Kaluza-Klein (KK) reduction, we will derive the ten dimensional
superfields which are necessary in constructing the superstring
action. In this way, we obtain another example of the Green-Schwarz
superstring action with $\kappa$-symmetry for a background with
non-maximal supersymmetry, whose supergravity constraints can be
completely solved.

The organization of this paper is as follows.  In the next section,
we review the approach of \cite{Kallosh:1998zx}, where one can read
off the 10d/11d superfields for a background described by
superalgebra and the associated coset structure. In section 3, the
superfields for the eleven dimensional plane wave background are
obtained starting from the symmetry superalgebra of the background.
The KK reduction is performed in Sec.~\ref{KKreduction}, which leads
to the ten dimensional superfields. In Sec.~\ref{stringaction}, the
complete type IIA superstring action on IIA plane wave background is
written down.  As a check, we show that the action reduces correctly
to the light-cone gauge fixed action constructed previously.
Sec.~\ref{concl} is devoted to conclusion. Our notation is
summarized in an appendix.

\section{Preliminary}
If underlying geometry can be described by superalgebra and
associated  supercoset structure, the supergravity constraints can
be completely solved. We review this approach by closely following
\cite{Kallosh:1998zx} and apply it to 11-dimensional plane wave
background in Sec. \ref{pp}.

The general superalgebra is of the form
\begin{eqnarray}
[B_{\hat{r}},
B_{\hat{s}}]&=&f_{\hat{r}\hat{s}}^{\,\,\,\,\hat{t}}B_{\hat{t}}
\nonumber  \\
\{F_{\hat{a}}, B_{\hat{r}}] &=&f_{\hat{a} \hat{r}}^{\,\,\,\,\hat{b}}F_{\hat{b}} \nonumber \\
 \{F_{\hat{a}}, F_{\hat{b}}\}&=&f_{\hat{a}
 \hat{b}}^{\,\,\,\,\hat{r}}F_{\hat{r}}
\end{eqnarray}
where $B_{\hat{r}}$ and $F_{\hat{a}}$ are the bosonic  and fermionic
generators, respectively and $f$ are the structure constants. The
differential operator is defined to be
\begin{equation}
D=d+L^{\hat{r}}B_{\hat{r}}+L^{\hat{a}}B_{\hat{a}}
\end{equation}
where $L^{\hat{r}}, L^{\hat{a}}$ are the left-invariant Cartan
one-forms. The equation $D^2$ leads to the usual Mauer-Cartan
equation
\begin{eqnarray}
dL^{\hat{r}}+\frac{1}{2}f_{\hat{s}\hat{t}}^{\,\,\,\,
\hat{r}}L^{\hat{s}} \wedge L^{\hat{t}}
-\frac{1}{2}f_{\hat{a}\hat{b}}^{\,\,\,\, \hat{r}}L^{\hat{a}}\wedge
L^{\hat{b}}&=&0 \nonumber
\\
dL^{\hat{a}}+f_{\hat{r}\hat{t}}^{\,\,\,\, \hat{a}}L^{\hat{r}}\wedge
L^{\hat{t}} &=&0
\end{eqnarray}
If the underlying space has the supercoset structrure, the
Mauer-Cartan equation can be solved completely. Writing
\begin{equation}
G=g(x)e^{\theta F} \,,
\end{equation}
we find that
\begin{equation}
G^{-1}dG=e^{-\theta F}De^{\theta F}
\end{equation}
with $D=d+L_0^{\hat{a}}B_{\hat{a}}$. Here $L_0^{\hat{a}}$ is defined
as
\begin{equation}
L^{\hat{A}}=L_0^{\hat{A}}(x)+\tilde{L}^{\hat{A}}(x, \theta)
\end{equation}
with $\hat{A}=(\hat{r}, \hat{a})$ collectively. If one introduces
$\lambda$ dependence by $\theta\rightarrow \lambda\theta$,
\begin{equation}
e^{-\lambda \theta}de^{\lambda \theta}=\tilde{L}_{\lambda}^{\hat{r}}
B_{\hat{r}}+\tilde{L}_{\lambda}^{\hat{a}}F_{\hat{a}} \,.
\end{equation}
By differentiating both sides and using the superalgebra, we obtain
\begin{eqnarray}
\partial_\lambda \tilde{L}_\lambda^{\hat{r}}
  &=&\theta^{\hat{a}}\tilde{L}_\lambda^{\hat{b}}
      f_{\hat{a}\hat{b}}^{\hat{r}}
\nonumber \\
\partial_\lambda \tilde{L}_\lambda^{\hat{a}}
  &=&d \theta^{\hat{a}}-\theta^{\hat{b}}
      f_{\hat{b} \hat{r}}^{\hat{a}} \tilde{L}_\lambda^{\hat{r}} \,.
\end{eqnarray}
With the initial condition
$\tilde{L}_{\lambda=0}^{\hat{r}}=\tilde{L}_{\lambda=0}^{\hat{a}}=0$,
these equations can be solved
\begin{eqnarray}
L^{\hat{a}} &=&
    \left(   \frac{\sinh {\mathcal M}}{{\mathcal M}}
    \right)^{\hat{a}}_{\hat{b}} d\theta^{\hat{b}} \nonumber \\
L^{\hat{r}}&=&
L^{\hat{r}}_0 + 2\theta^{\hat{a}} f_{\hat{a}\hat{b}}^{\,\,\,\,\hat{r}}
  \left(
       \frac{\sinh^2 {\mathcal M}/2}{{\mathcal M}^2}
  \right)^{\hat{b}}_{\hat{r}}
   d\theta^{\hat{r}} \,,
\end{eqnarray}
where $({\mathcal M}^2)^{\hat{a}}_{\hat{b}} = - \theta^{\hat{c}}
f_{\hat{c}\hat{r}}{}^{\hat{a}} \theta^{\hat{d}}
f_{\hat{d}\hat{b}}{}^{\hat{r}}$.
If we choose $\theta$ to be the standard fermion coordinates of the
superspace, this gives the superspace geometry in Wess-Zumino gauge.

If the above background describes the 11-d background, the membrane
action can be written as
\begin{equation}
S=\int d^3\xi \sqrt{-det L^{\hat{r}}_{\hat{i}}
L^{\hat{s}}_{\hat{j}}\eta_{\hat{r}\hat{s}}}-\frac{1}{6}L^{\hat{A}}_{\hat{i}}L^{\hat{B}}_{\hat{j}}
L^{\hat{C}}_{\hat{k}} B_{\hat{A}\hat{B}\hat{C}}
\end{equation}
where $B_{\hat{A}\hat{B}\hat{C}}$ represents 3-form superfield,
which should be determined separately. Later we will do this for
11-d plane-wave background in section \ref{pp}.

Once we obtain the membrane action in 11-d then we carry out the
double dimensional reduction to obtain the string action in 10-d
plane wave background.

\section{Superfields for the eleven dimensional plane wave background}\label{pp}

\subsection{Symmetry superalgebra}

The eleven dimensional plane wave background
\cite{KowalskiGlikman:1984wv} is one  of the maximally
supersymmetric solutions in eleven dimensional supergravity
\cite{maxsol} and is given by
\begin{align}
ds^2 &= 2 dx^+ dx^- -
 \left( \sum_{\hat{i}=1}^{3} \frac{\mu^2}{9} (x^{\hat{i}})^2
  + \sum_{\hat{i}'=4}^{9} \frac{\mu^2}{36} (x^{\hat{i}'})^2
 \right) (dx^-)^2 + (dx^{\hat{I}})^2 \, ,
\notag \\
F_{-123} &= \mu \,,
\label{pp-wave}
\end{align}
where $x^\pm = \frac{1}{\sqrt{2}} (x^{11} \pm x^0)$ and
$\hat{I} = (\hat{i},\hat{i}')$.  Apart from an interesting solution,
this background has an important connection with the AdS type
backgrounds such as AdS$_4 \times S^7$ or AdS$_7 \times S^4$ through
the Penrose limit \cite{Blau:2002dy}.  In this case, the dimensionful
parameter $\mu$ is inversely proportional to the radius $R$ of the AdS
space, $\mu \propto 1/R$.

More detailed information on the plane wave background
(\ref{pp-wave}) is given by the underlying superalgebra, which has
been obtained from the investigation of isometries in
\cite{FigueroaO'Farrill:2001nz}. Regarding the relation between the
plane wave and the AdS type backgrounds, it has been shown that the
same symmetry superalgebra can be derived also from the superalgebra
of AdS$_4 \times S^7$ or AdS$_7 \times S^4$ via the
In\"{o}n\"{u}-Wigner contraction which can be regarded as the
algebraic version of the Penrose limit \cite{Hatsuda:2002kx}.
Referring to Refs.~\cite{FigueroaO'Farrill:2001nz,Hatsuda:2002kx},
the superalgebra is given as follows.\footnote{We follow the
algebra given in \cite{FigueroaO'Farrill:2001nz} but with the
notation of \cite{Hatsuda:2002kx} for generators.  We also take a
rescaling of supercharge as $Q \rightarrow Q/\sqrt{2}$ for our
convenience.} Firstly, the commutation relations between the bosonic
generators are
\begin{align}
& [ P_{\hat{I}}, P_- ] = - P^*_{\hat{I}} \,, \quad
[ P^*_{\hat{i}}, P_- ] = \frac{\mu^2}{9} P_{\hat{i}} \,, \quad
[ P^*_{\hat{i'}}, P_- ] = \frac{\mu^2}{36} P_{\hat{i'}} \,,
\notag \\
& [ P^*_{\hat{i}}, P_{\hat{j}} ]
  = - \frac{\mu^2}{9} \eta_{\hat{i}\hat{j}} P_+ \,, \quad
[ P^*_{\hat{i'}}, P_{\hat{j'}} ]
  = - \frac{\mu^2}{36} \eta_{\hat{i'}\hat{j'}} P_+ \,,
\notag \\
& [ J_{\hat{I}\hat{J}}, P_{\hat{K}} ]
  = 2 \eta_{\hat{J}\hat{K}} P_{\hat{I}} \,, \quad
[ J_{\hat{I}\hat{J}}, P^*_{\hat{K}} ]
  = 2 \eta_{\hat{J}\hat{K}} P^*_{\hat{I}} \,, \quad
[ J_{\hat{I}\hat{J}}, J_{\hat{K}\hat{L}} ]
  = 4 \eta_{\hat{J}\hat{K}} J_{\hat{I}\hat{L}} \,,
\label{bb-alg}
\end{align}
where $\hat{i}=1 \cdots 3, \hat{i}'=4 \cdots 9$ and
$\hat{I},\hat{J}$ denote  $SO(9)$ vector index. The details of the
convention are summarized at the appendix. Also $P$ and $J$ denote
the translation and the rotation generators respectively and
\begin{align}
P_\pm \equiv \frac{1}{\sqrt{2}}(P_{11} \pm P_0) \,, \quad
P^*_{\hat{i}} \equiv J_{\hat{i}0} \,, \quad
P^*_{\hat{i}'} \equiv J_{\hat{i}'11} \,.
\label{pdef}
\end{align}
Secondly, the algebra between the bosonic and the fermionic generators
is\footnote{Throughout the paper, we suppress the spinor indices
unless there is some confusion.}
\begin{align}
& [ P_-, Q_+ ] = -\frac{\mu}{4} Q_+ \Pi \,, \quad
[ P_-, Q_- ] = -\frac{\mu}{12} Q_- \Pi \,, \quad
[ P_{\hat{i}}, Q_- ]
  = \frac{\mu}{6} Q_+ \Gamma^- \Pi \Gamma_{\hat{i}} \,,
\notag \\
& [ P_{\hat{i}'}, Q_- ]
  = \frac{\mu}{12} Q_+ \Gamma^- \Pi \Gamma_{\hat{i}} \,, \quad
[ P^*_{\hat{i}}, Q_- ]
  = \frac{\mu^2}{18} Q_+ \Gamma_{\hat{i}} \Gamma^- \,, \quad
[ P^*_{\hat{i}}, Q_- ]
  = \frac{\mu^2}{72} Q_+ \Gamma_{\hat{i}'} \Gamma^- \,,
\label{bf-alg}
\end{align}
where $\Gamma$'s are $32 \times 32$ Dirac gamma matrices,
\begin{equation}
\Pi \equiv \Gamma^{123} \,,
\end{equation}
and the supersymmetry generator $Q$ with 32 components has been split
into two parts by introducing a projection operator ${\mathcal P}_\pm$ as
\begin{equation}
{\mathcal P}_\pm \equiv \frac{1}{2} \Gamma_\pm \Gamma_\mp \,, \quad
Q_\pm \equiv Q {\mathcal P}_\pm \,, \quad
Q = Q_+ + Q_- \,.
\label{proj}
\end{equation}
Finally, the algebra of supercharges is
\begin{align}
& \{ Q_+, Q_+ \} = - 2C \Gamma^+ P_+ \,,
\notag \\
& \{ Q_-, Q_- \}
  = - 2C \Gamma^- P_-
    - \frac{\mu}{3} C \Gamma^- \Pi
         \Gamma^{\hat{i}\hat{j}} J_{\hat{i}\hat{j}}
    + \frac{\mu}{6} C \Gamma^- \Pi
         \Gamma^{\hat{i}'\hat{j}'} J_{\hat{i}'\hat{j}'} \,,
\notag \\
& \{ Q_+, Q_- \}
  = - 2C \Gamma^{\hat{I}} P_{\hat{I}}
    - \frac{6}{\mu} C \Pi \Gamma^{\hat{i}} P^*_{\hat{i}}
    - \frac{12}{\mu} C \Pi \Gamma^{\hat{i}'} P^*_{\hat{i}'} \,,
\label{ff-alg}
\end{align}
where $C$ is the charge conjugation matrix satisfying $C
\Gamma^{\hat{r}} C^{-1} = - \Gamma^{\hat{r} T}$.  Here, it seems
that the last anticommutation relation is problematic in the limit
of $\mu \rightarrow 0$, which is nothing but the flat spacetime
limit as can be seen from the background (\ref{pp-wave}).  Although
it should be reduced to the superalgebra of flat spacetime in this
limit, it is apparently divergent.  However this is just an artifact
of convention. Indeed, as pointed out explicitly in
\cite{Hatsuda:2002kx}, a proper rescaling of superalgebra generators
cures the problem of flat spacetime limit.

\subsection{Cartan one-forms}

The symmetry superalgebra, (\ref{bb-alg}), (\ref{bf-alg}),
and (\ref{ff-alg}), is the full superalgebra of the
plane wave background (\ref{pp-wave}) and one feature of it is that it
has the form of the algebraic structure of coset superspace
${\mathcal G}/{\mathcal H}$\cite{FigueroaO'Farrill:2001nz}.
(The whole generators are those of the group ${\mathcal G}$ and the
rotation generators correspond to those of the stability subgroup
${\mathcal H}$.)  This implies that, by starting from the full superalgebra
and following the general prescription suggested
in \cite{Metsaev:1998it,Kallosh:1998zx}, we can
read off the superspace geometry represented by superfields which is
our goal in this section.

In the case where the superspace has the coset structure, the superfields
are expressed in terms of the left-invariant Cartan one-forms.  If we
let $G$ be the coset representative, then
\begin{equation}
G^{-1} d G =
L^{\hat{r}} P_{\hat{r}}
+ L^{\hat{i}}_* P^*_{\hat{i}}
+ L^{\hat{i}'}_*P^*_{\hat{i}'}
+ \frac{1}{2} L^{\hat{I}\hat{J}} J_{\hat{I}\hat{J}}
+ {\mathbf L}^+Q_+ + {\mathbf L}^-Q_-
\label{c1form}
\end{equation}
gives the left-invariant Cartan one-forms, $L^{\hat{r}}$,
$L^{\hat{i}}_*$, $L^{\hat{i}'}_*$,
$L^{\hat{I}\hat{J}}$, ${\mathbf L}^+$, and ${\mathbf L}^-$, where
the spinorial one-forms ${\mathbf L}^\pm$ are the projections of
the spinorial one-form $L$ just like $Q_\pm$ of Eq.~(\ref{proj});
\begin{align}
{\mathbf L}^\pm \equiv {\mathcal P}_\pm L \,, \quad
L = {\mathbf L}^+ + {\mathbf L}^- \,.
\end{align}
We note that $L^{\hat{i}\hat{j}'}=0$ because
there is no generator $J_{\hat{i}\hat{j}'}$ in the superalgebra of
(\ref{bb-alg}), (\ref{bf-alg}), and (\ref{ff-alg}).
Now, from the integrability of (\ref{c1form}) and the superalgebra,
the left-invariant Cartan one-forms turn out to satisfy the
Maurer-Cartan equations,
\begin{align}
& dL^+ - \frac{\mu^2}{9} L^{\hat{i}}_* \wedge L^{\hat{i}}
 - \frac{\mu^2}{36} L^{\hat{i}'}_* \wedge L^{\hat{i}'}
 + \bar{L} \wedge \Gamma^+ L = 0 \,,
\notag \\
& dL^- + \bar{L} \wedge \Gamma^- L =0 \,,
\notag \\
& dL^{\hat{i}} + \frac{\mu^2}{9} L^{\hat{i}}_* \wedge L^-
 + L^{\hat{i}\hat{j}} \wedge L^{\hat{j}}
 + \bar{L} \wedge \Gamma^{\hat{i}} L = 0 \,,
\notag \\
& dL^{\hat{i}'} + \frac{\mu^2}{36} L^{\hat{i}'}_* \wedge L^-
 + L^{\hat{i}' \hat{j}'} \wedge L^{\hat{j}'}
 + \bar{L} \wedge \Gamma^{\hat{i}'} L = 0 \,,
\notag \\
& dL^{\hat{i}}_* - L^{\hat{i}} \wedge L^-
 + L^{\hat{i}\hat{j}} \wedge L^{\hat{j}}_*
 + \frac{3}{\mu} \bar{L} \wedge \Pi \Gamma^{\hat{i}} L = 0 \,,
\notag \\
& dL^{\hat{i}'}_* - L^{\hat{i}'} \wedge L^-
 + L^{\hat{i}' \hat{j}'} \wedge L^{\hat{j}'}_*
 + \frac{6}{\mu} \bar{L} \wedge \Pi \Gamma^{\hat{i}'} L = 0 \,,
\notag \\
& dL^{\hat{i}\hat{j}} + L^{\hat{k}\hat{i}} \wedge L^{\hat{j}\hat{k}}
 + \frac{\mu}{6} \bar{L} \wedge \Gamma^- \Pi \Gamma^{\hat{i}\hat{j}} L
 = 0 \,,
\notag \\
& dL^{\hat{i}' \hat{j}'}
 + L^{\hat{k}' \hat{i}'} \wedge L^{\hat{j}' \hat{k}'}
 + \frac{\mu}{12} \bar{L} \wedge \Gamma^- \Pi \Gamma^{\hat{i}' \hat{j}'} L
 = 0 \,,
\notag \\
& dL
 + \frac{1}{4} L^{\hat{I}\hat{J}} \wedge \Gamma_{\hat{I}\hat{J}} L
 - \frac{\mu}{12} L^- \wedge \Pi ( \Gamma^- \Gamma^+ + 1 ) L
 + \frac{\mu}{6} L^{\hat{i}} \wedge \Gamma^- \Pi \Gamma_{\hat{i}} L
\notag \\
& + \frac{\mu}{12} L^{\hat{i}'} \wedge \Gamma^- \Pi \Gamma_{\hat{i}'} L
 + \frac{\mu^2}{18} L^{\hat{i}}_* \wedge \Gamma_{\hat{i}} \Gamma^- L
 + \frac{\mu^2}{72} L^{\hat{i}'}_* \wedge \Gamma_{\hat{i}'} \Gamma^- L
 =0 \,,
\label{mceq}
\end{align}
where $\bar{L} = L^T C$.

\subsection{Eleven dimensional superfields}

The left-invariant Cartan one-forms are functions of the supercoordinate
composed of $x^{\hat{r}}$ and $\theta^{\hat{a}}$ parametrizing the
superspace, and each of them has its own expansion in terms of $\theta$.
Since their expanded form is eventually necessary in the construction of
the superstring action, we now determine them to all orders in $\theta$.
In order to do this, we begin with making a
particular choice of the coset representative $G$ in (\ref{c1form})
as
\begin{equation}
G(x,\theta) = g(x) e^{\theta^+ Q_+ + \theta^- Q_-} \,,
\label{wzgauge}
\end{equation}
which is known as the Wess-Zumino type parametrization.  The bosonic
factor $g(x)$ is for the purely bosonic part of the superspace and is
left unspecified. The fermionic coordinates in the exponential factor
are defined by, using the projection operator in (\ref{proj}),
\begin{gather}
\theta^\pm \equiv {\mathcal P}_\pm \theta \,, \quad
\theta = \theta^+ + \theta^- \,.
\end{gather}

As a next step, we take the rescaling $\theta \rightarrow \lambda \theta$
with an auxiliary parameter $\lambda$ \cite{Metsaev:1998it,Kallosh:1998zx}
and put a subscript $\lambda$ for rescaled quantities such as, for example,
$G_\lambda = G(x, \lambda \theta)$ and
$L^{\hat{r}}_\lambda = L^{\hat{r}} (x, \lambda \theta)$.
Then the differentiation with respect to $\lambda$ of (\ref{c1form})
with (\ref{wzgauge}) and the superalgebra,
(\ref{bb-alg}), (\ref{bf-alg}), and (\ref{ff-alg}), lead us to have
\begin{align}
& \partial_\lambda L^{\hat{r}}_\lambda
  = 2 \bar{L}_\lambda \Gamma^{\hat{r}} \theta \,,
\notag \\
& \partial_\lambda L^{\hat{i}}_{*\lambda} =
   \frac{6}{\mu} \bar{L}_\lambda \Pi \Gamma^{\hat{i}} \theta \,, \quad
\partial_\lambda L^{\hat{i}'}_{*\lambda} =
   \frac{12}{\mu} \bar{L}_\lambda \Pi \Gamma^{\hat{i}'} \theta \,,
\notag \\
& \partial_\lambda L^{\hat{i}\hat{j}}_\lambda =
   \frac{\mu}{3} \bar{L}_\lambda \Gamma_+ \Pi \Gamma^{\hat{i}\hat{j}}
    \theta \,,
 \quad
\partial_\lambda L^{\hat{i}'\hat{j}'}_\lambda =
  -\frac{\mu}{6} \bar{L}_\lambda \Gamma_+ \Pi \Gamma^{\hat{i}'\hat{j}'}
   \theta \,,
\notag \\
& \partial_\lambda L_\lambda =
  d \theta + \frac{1}{4} L^{\hat{I}\hat{J}}_\lambda \Gamma_{\hat{I}\hat{J}} \theta
  - \frac{\mu}{12} L^-_\lambda \Pi (\Gamma^- \Gamma^+ +1 ) \theta
\notag \\
  & + \frac{\mu}{6}
      \left(L^{\hat{i}}_\lambda \Gamma^- \Pi \Gamma_{\hat{i}}
          +\frac{1}{2}L^{\hat{i}'}_\lambda \Gamma^- \Pi \Gamma_{\hat{i}'}
          +\frac{\mu}{3} L^{\hat{i}}_{*\lambda} \Gamma_{\hat{i}}\Gamma^-
           +\frac{\mu}{12} L^{\hat{i}'}_{*\lambda} \Gamma_{\hat{i}'}
             \Gamma^-
      \right)\theta \,.
\label{teq}
\end{align}
As pointed out in \cite{Kallosh:1998zx}, these first-order differential
equations have the structure of coupled harmonic oscillators, and thus
can be solved exactly.  To solve these equations, we first impose
the initial conditions for the Cartan one-forms as
\begin{align}
L^{\hat{r}}_{\lambda=0} = \hat{e}^{\hat{r}} \,, \quad
L^{\hat{i}\hat{j}}_{\lambda=0} = \hat{\omega}^{\hat{i}\hat{j}} \,, \quad
L^{\hat{i}'\hat{j}'}_{\lambda=0} = \hat{\omega}^{\hat{i}'\hat{j}'} \,,
 \quad
L_{\lambda=0} = 0 \,,
\label{ini1}
\end{align}
where $\hat{e}^{\hat{r}}$ is the elfbein and
$\hat{\omega}^{\hat{i}\hat{j}}$, $\hat{\omega}^{\hat{i}'\hat{j}'}$
are the spin connections of the plane wave geometry. We note that
$\hat{\omega}^{\hat{i}\hat{j}'}=0$ automatically due to the fact
that $L^{\hat{i}\hat{j}'}=0$ as alluded to previously.  As for the
remaining Cartan one-forms, one may be tempted to take
$L^{\hat{i}}_{*\lambda=0} = \hat{\omega}^{\hat{i}0}$ and
$L^{\hat{i}'}_{*\lambda=0} = \hat{\omega}^{\hat{i}'11}$ from the
definition of (\ref{pdef}).  However, this is naive expectation.
Indeed, if we compare the purely bosonic part of the Maurer-Cartan
equation (\ref{mceq}) with the usual Cartan structure equation, we
can see that the correct initial conditions are
\begin{equation}
L^{\hat{i}}_{*\lambda=0} = - \frac{9}{\mu^2} \hat{\omega}^{+\hat{i}}
  \,,  \quad
L^{\hat{i}'}_{*\lambda=0} = - \frac{36}{\mu^2} \hat{\omega}^{+\hat{i}'} \,,
\label{ini2}
\end{equation}
together with a consistency condition $\hat{\omega}^{-\hat{r}}=0$.
The initial conditions, (\ref{ini1}) and (\ref{ini2}), form an enough set
of data for solving the differential equations of (\ref{teq}).
Furthermore, they give us an information about non-vanishing
derivative at $\lambda=0$ \cite{deWit:1998yu}, which is identified with the
eleven dimensional covariant derivative for the fermionic coordinate
denoted by
$\widehat{D}\theta$;
\begin{align}
\widehat{D} \theta \equiv \partial_\lambda L_\lambda |_{\lambda=0} \,.
\end{align}
If we consider the covariant derivatives for $\theta^+$ and $\theta^-$
separately, their explicit expressions are obtained as
\begin{eqnarray}
\widehat{D} \theta^+
 &=& d \theta^+
  + \frac{1}{4} \hat{\omega}^{\hat{I}\hat{J}} \Gamma_{\hat{I}\hat{J}}
  \theta^+
  + \frac{1}{2} \hat{\omega}^{+\hat{I}} \Gamma_{+\hat{I}}\theta^-
  - \frac{\mu}{4} \hat{e}^- \Pi \theta^+
\notag \\
 & & + \frac{\mu}{6}
     \left(
       \hat{e}^{\hat{i}} \Gamma^- \Pi \Gamma_{\hat{i}}
       +\frac{1}{2} \hat{e}^{\hat{i}'} \Gamma^- \Pi \Gamma_{\hat{i}'}
     \right) \theta^- \,,
\notag \\
\widehat{D} \theta^-
 &=& d \theta^-
  + \frac{1}{4} \hat{\omega}^{\hat{I}\hat{J}} \Gamma_{\hat{I}\hat{J}}
  \theta^-
  - \frac{\mu}{12} \hat{e}^- \Pi \theta^-  \,.
\label{11cov}
\end{eqnarray}

Now it is straightforward to solve the equations (\ref{teq}) with
the initial conditions (\ref{ini1}) and (\ref{ini2}).  After
setting $\lambda=1$, that is, $L = L_{\lambda=1}$, we finally have
\begin{align}
L^{\hat{r}} &= \hat{e}^{\hat{r}}
 -2  \sum^{15}_{n=0} \frac{1}{(2n+2)!} \bar{\theta}\Gamma^{\hat{r}}
   {\mathcal M}^{2n} \widehat{D} \theta \,,
\notag \\
L_*^{\hat{i}} &= - \frac{9}{\mu^2} \hat{\omega}^{+\hat{i}}
 - \frac{6}{\mu} \sum^{15}_{n=0} \frac{1}{(2n+2)!}
  \bar{\theta} \Pi \Gamma^{\hat{i}}
  {\mathcal M}^{2n} \widehat{D} \theta \,,
\notag \\
L_*^{\hat{i}'} &= - \frac{36}{\mu^2} \hat{\omega}^{+\hat{i}'}
 - \frac{12}{\mu} \sum^{15}_{n=0} \frac{1}{(2n+2)!}
  \bar{\theta} \Pi \Gamma^{\hat{i}'}
  {\mathcal M}^{2n} \widehat{D} \theta \,,
\notag \\
L^{\hat{i}\hat{j}} &= \hat{\omega}^{\hat{i}\hat{j}}
 - \frac{\mu}{3} \sum^{15}_{n=0} \frac{1}{(2n+2)!}
  \bar{\theta} \Gamma^- \Pi \Gamma^{\hat{i}\hat{j}}
  {\mathcal M}^{2n} \widehat{D} \theta \,,
\notag \\
L^{\hat{i}' \hat{j}'} &= \hat{\omega}^{\hat{i}'\hat{j}'}
 + \frac{\mu}{6} \sum^{15}_{n=0} \frac{1}{(2n+2)!}
  \bar{\theta} \Gamma^- \Pi \Gamma^{\hat{i}' \hat{j}'}
  {\mathcal M}^{2n} \widehat{D} \theta \,,
\notag \\
L &= \sum^{16}_{n=0} \frac{1}{(2n+1)!} {\mathcal M}^{2n}
    \widehat{D} \theta \,,
\label{11super}
\end{align}
where $\widehat{D} \theta = \widehat{D} \theta^+ +
\widehat{D} \theta^-$ and ${\mathcal M}^2$ is the $32 \times 32$
matrix given by
\begin{align}
{\mathcal M}^2 = \frac{\mu}{6}
 \bigg[
&
   (\Pi (\Gamma^- \Gamma^+ +1 ) \theta)
   (\bar{\theta} \Gamma^- {\mathcal P}_- )
 -
   2 (\Gamma^- \Pi \Gamma_{\hat{i}} {\mathcal P}_- \theta)
   (\bar{\theta} \Gamma^{\hat{i}} )
 -
   (\Gamma^- \Pi \Gamma_{\hat{i}'} {\mathcal P}_- \theta)
   (\bar{\theta} \Gamma^{\hat{i}'} )
\notag \\
& -
   2 (\Gamma_{\hat{i}} \Gamma^- {\mathcal P}_- \theta)
   (\bar{\theta} \Pi \Gamma^{\hat{i}} )
 -
   (\Gamma_{\hat{i}'} \Gamma^- {\mathcal P}_- \theta)
   (\bar{\theta} \Pi \Gamma^{\hat{i}'} )
 -
   \frac{1}{2} (\Gamma_{\hat{i}\hat{j}} \theta)
   (\bar{\theta} \Gamma^- \Pi \Gamma^{\hat{i}\hat{j}}
       {\mathcal P}_-)
\notag \\
& +
   \frac{1}{4} (\Gamma_{\hat{i}'\hat{j}'} \theta)
   (\bar{\theta} \Gamma^- \Pi \Gamma^{\hat{i}'\hat{j}'}
       {\mathcal P}_-)
 \bigg] \,.
\label{m2mat}
\end{align}

The left-invariant Cartan one-forms of (\ref{11super}) are the
superfields describing the superspace geometry.  On the other hand,
there is one more ingredient in the superspace for the eleven
dimensional plane wave background. It is the three-form superfield
$\widehat{B}$, which forms the Wess-Zumino part of the supermembrane
action.  Basically, the problem is to find the closed four-form
superfield $H$ from a certain combination of various products of the
superfields (\ref{11super}) and relate it to $\widehat{B}$ through
the local equation $d\widehat{B}=H$.  In the present case, there are
two possible candidates for $H$, which are $\bar{L} \wedge
\Gamma_{\hat{r}\hat{s}} L \wedge L^{\hat{r}} \wedge L^{\hat{s}}$ and
$L^{\hat{r}} \wedge L^{\hat{s}} \wedge L^{\hat{t}} \wedge
L^{\hat{u}} F_{\hat{r}\hat{s}\hat{t}\hat{u}}$.\footnote{We note that
two terms are the same with those for the supermembrane in AdS$_4
\times S^7$ or AdS$_7 \times S^4$ \cite{deWit:1998yu}.  We may think
that this is natural since the plane wave background is the Penrose
limit of these two AdS type backgrounds and hence the formal
structure of the Wess-Zumino part is expected to be unchanged
although the details are different.  For more comprehensive study on
the Wess-Zumino part of the supermembrane as well as the super
fivebrane in various backgrounds, see for example
Ref.~\cite{Sakaguchi:2006pg} where the systematic
Chevalley-Eilenberg cohomology\cite{DeAzcarraga:1989vh} has been
used.} We take a linear combination of these terms for $H$ and fix
the relative coefficient by requiring the closedness of $H$, that
is, $dH=0$.  This process is performed by using the Maurer-Cartan
equation (\ref{mceq}) and the eleven dimensional Fierz identity
$(C\Gamma_{\hat{r}\hat{s}})_{(\hat{a}\hat{b}}
(C\Gamma^{\hat{s}})_{\hat{c}\hat{d})} = 0$, and the resulting
expression is obtained as
\begin{equation}
H = \frac{1}{4!} ( L^{\hat{r}} \wedge L^{\hat{s}}
\wedge L^{\hat{t}} \wedge L^{\hat{u}} F_{\hat{r}\hat{s}\hat{t}\hat{u}}
+
12 \bar{L} \wedge \Gamma_{\hat{r}\hat{s}} L \wedge L^{\hat{r}}
\wedge L^{\hat{s}} ) \,,
\end{equation}
where the overall multiplicative constant has been fixed such that it
leads to the standard form of the Wess-Zumino term, for example as
in \cite{deWit:1998yu}.

Having the expression of $H$, we first apply the trick of rescaling
$\theta \rightarrow \lambda \theta$ to $H$;
$H_\lambda = H(x, \lambda\theta)$.
Then the following identity provides the equation for finding
$\widehat{B}$.
\begin{equation}
H_{\lambda=1} = H_{\lambda=0} + \int^1_0 \! d\lambda \,
  \partial_\lambda H_\lambda \,.
\label{ht}
\end{equation}
By using the differential equations (\ref{teq}), we find that
\begin{equation}
\partial_\lambda H_\lambda = d (\bar{\theta} \Gamma_{\hat{r}\hat{s}}
    L_\lambda \wedge L^{\hat{r}}_\lambda \wedge L^{\hat{s}}_\lambda ) \,.
\end{equation}
If we plug this into (\ref{ht}) and use the local relation
$d \widehat{B} = H = H_{\lambda=1}$, then we obtain
\begin{equation}
\widehat{B}
= \frac{1}{6} \hat{e}^{\hat{r}} \wedge \hat{e}^{\hat{s}} \wedge
    \hat{e}^{\hat{t}} \widehat{C}_{\hat{r}\hat{s}\hat{t}}
 + \int^1_0 d\lambda \; \bar{\theta} \Gamma_{\hat{r}\hat{s}}
    L_\lambda \wedge L^{\hat{r}}_\lambda \wedge L^{\hat{s}}_\lambda \,,
\label{s3f}
\end{equation}
where the first term on the right hand side has resulted from
$H_{\lambda=0}$ and $\widehat{C}_{\hat{r}\hat{s}\hat{t}}$ is three-form
gauge field whose field strength is
$\widehat{F}_{\hat{r}\hat{s}\hat{t}\hat{u}} =
4 \partial_{[ \hat{r}} \widehat{C}_{\hat{s}\hat{t}\hat{u} ] }$.

\section{Kaluza-Klein reduction}
\label{KKreduction}

Having the eleven dimensional superfields for the eleven dimensional
plane wave background in hand, we carry out the KK reduction along a
spatial isometry direction of the plane wave geometry and obtain the
superfields in ten dimensions.  Since the Cartan one-forms,
$L^{\hat{r}}$ and $L$, and the three-form superfield $\widehat{B}$
are the necessary elements for constructing the IIA superstring
action, we will focus only on them.

\subsection{Ten dimensional bosonic geometry}
\label{10dbosgeo}

As we can see from the eleven dimensional plane wave geometry
(\ref{pp-wave}), there is no explicit spatial isometry direction.
However, it has been shown \cite{Sugiyama:2002tf,Hyun:2002wu} that the
geometry can be made to have such an isometry under a suitable
coordinate transformation, which is taken by following the prescription suggested in \cite{Michelson:2002wa}. In our convention, if $x^9$
is taken to be the desired isometry direction, the transformation is
given by
$x^+ \rightarrow x^+ - (\mu/6) x^4 x^9 $,
$x^4 \rightarrow \cos (\mu x^-/6) x^4 - \sin(\mu x^-/6) x^9$, and
$x^9 \rightarrow \sin (\mu x^-/6) x^4 + \cos(\mu x^-/6) x^9$
without changing other coordinates.
From the transformed geometry, we may choose the elfbein as
\begin{equation}
\hat{e}^+ = dx^+ - \frac{1}{2} A (x^I) dx^- \,, \quad
\hat{e}^- = dx^- \,, \quad
\hat{e}^I = dx^I \,, \quad
\hat{e}^9 = dx^9 + \frac{\mu}{3} x^4 dx^- \,,
\label{elf}
\end{equation}
where
\begin{equation}
A(x^I) =
 \sum_{i=1}^{4} \frac{\mu^2}{9} (x^i)^2
  + \sum_{i'=5}^{8} \frac{\mu^2}{36} (x^{i'})^2 \,,
\label{axi}
\end{equation}
and $I = (i,i')= 1, \dots, 8$.  Clearly, this choice has a suitable form
for the KK reduction along $x^9$ basically because it satisfies
the KK ansatz
\begin{align}
\hat{e}^r_9 = 0 \,.
\label{kkcond}
\end{align}

The above elfbein (\ref{elf}) is taken to be a parametrization of the
purely bosonic part of the super elfbein, that is, the Cartan one-form
$L^{\hat{r}}$ of (\ref{11super}).  For its validity, we would like to
note that it satisfies the two consistency conditions of the last
section for the spin connection stemming from the symmetry superalgebra,
which are $\hat{\omega}^{\hat{i}\hat{j}'}=0$ and
$\hat{\omega}^{-\hat{r}}=0$ (just below of Eqs.~(\ref{ini1}) and
(\ref{ini2})).  Indeed, through the actual computation
of the spin connection with the elfbein (\ref{elf}), we see
that\footnote{We note that $\hat{\omega}^{49}$ is the type
of $\hat{\omega}^{\hat{i}'\hat{j}'}$.}

\begin{align}
\hat{\omega}^{+I} = -\frac{1}{2} \partial_I A dx^-
  + \frac{1}{2} \delta^{I4}
     \left(
        \frac{\mu^2}{9} x^4 dx^- + \frac{\mu}{6} dx^9
     \right) \,, \quad
\hat{\omega}^{+9} = \frac{\mu}{6} dx^4 \,, \quad
\hat{\omega}^{49} = -\frac{\mu}{6} dx^- \,,
\end{align}
and thus our parametrization is valid one.

Following the standard rule of KK reduction, we can directly
read off from the elfbein (\ref{elf}) the ten dimensional
quantities, that is, the zehnbein, the dilaton, and the
Ramond-Ramond (R-R) one-form gauge field.
In the string frame, the zehnbein is
\begin{align}
e^+ = dx^+ - \frac{1}{2} A (x^I) dx^- \,, \quad
e^- = dx^- \,, \quad
e^I = dx^I \,, \quad
\label{zehn}
\end{align}
from which the corresponding non-vanishing spin connection is obtained
as
\begin{equation}
\omega^{+I} = -\frac{1}{2} \partial_I A dx^- \,.
\label{10spin}
\end{equation}
The dilaton is trivially zero, $\phi=0$, and the field strength of
the R-R one-form gauge field is
\begin{align}
F_{-4} = -\frac{\mu}{3} \,.
\label{rr2}
\end{align}
Together with the R-R four-form field strength $F_{-123}=\mu$ descending
from the eleven dimensional plane wave background (\ref{pp-wave}),
(\ref{zehn}) and (\ref{rr2}) form the ten dimensional IIA plane wave
background.

As shown in \cite{Hyun:2002wu,Bena:2002kq} explicitly, the IIA plane
wave background is not maximally supersymmetric and preserves 24
supersymmetries.  To see this, let us consider the Killing spinor
equation for our ten dimensional background,\footnote{The Killing
spinor equation is obtained from the supersymmetry variation
of the dilatino field in the type IIA supergravity.  Although there is
another Killing spinor equation from the variation of the gravitino, it
is not necessary in the current discussion. For more details, see
\cite{Hyun:2002wu} for example.}
\begin{align}
\Gamma^- \Gamma^4 (1-\Gamma^{12349}) \epsilon = 0 \,.
\end{align}
If we write the 32 component $\epsilon$ as
$\epsilon = \epsilon^+ + \epsilon^-$ where
$\epsilon^\pm \equiv {\mathcal P}_\pm \epsilon$ with
the ${\mathcal P}_\pm$ of (\ref{proj}), and introduce a new projection
operator defined by
\begin{equation}
h_\pm \equiv \frac{1}{2} ( 1 \pm \Gamma^{12349} ) \,,
\label{hproj}
\end{equation}
then it is easy to see that $\epsilon^+$ and $h_+ \epsilon^-$ satisfy
the Killing spinor equation and correspond to the supersymmetry of the
IIA plane wave background.  Because two projection operators commute
with each other and each of them plays the role of filtering out half
the components of $\epsilon$, $\epsilon^+$ and $h_+ \epsilon^-$ have 16
and 8 independent components respectively.  This means that we get 24
supersymmetries in total.  The remaining 8 components represented by
$h_- \epsilon^-$ correspond to the broken supersymmetry.  If we state
this a little bit more, the projection operator
\begin{align}
h_- {\mathcal P}_-
\label{8proj}
\end{align}
allows us to pick out the components of the spinorial quantity
corresponding to the 8 broken supersymmetries.

\subsection{Ten dimensional superfields}

Now we turn to the KK reduction of eleven dimensional superfields.
Similar to the previous bosonic case, there is a condition (the
KK ansatz) that the super elfbein $L^{\hat{r}}$ should satisfy for the
consistent KK reduction.  It is $L^r_9=0$ regarded as the superspace
extension of (\ref{kkcond}) \cite{Duff:1987bx}.  To check this condition,
let us rewrite the eleven dimensional super-covariant derivative
$\widehat{D}\theta$ appearing in $L^{\hat{r}}$ of (\ref{11super}) in terms
of the ten dimensional quantities of the previous subsection.
Then, from the expression of $\widehat{D} \theta$ given in (\ref{11cov}),
we have
\begin{align}
\widehat{D}\theta
 &= \widehat{D}\theta^+ + \widehat{D}\theta^-
\notag \\
 &= D \theta + \frac{\mu}{6} \Gamma^{-4} h_- \theta^- \hat{e}^9 \,,
\label{11-10cov}
\end{align}
where $h_-$ is the projection operator defined in (\ref{hproj}) and
$D \theta$ is the ten dimensional super-covariant derivative identified as
\begin{eqnarray}
D \theta
&=& d \theta + \frac{1}{2} \omega^{+I} \Gamma_{+I} \theta
 + \frac{\mu}{12} ( 2 e^i \Gamma^- \Pi \Gamma_i +
     e^{i'} \Gamma^- \Pi \Gamma_{i'}
     - 2 e^4 \Gamma^- \Pi \Gamma_4 h_- ) \theta
\notag \\
& &  - \frac{\mu}{12} e^- \Pi ( \Gamma^- \Gamma^+ + 2h_- )\theta \,.
\label{10cov}
\end{eqnarray}
From (\ref{11-10cov}) and the expressions of ${\mathcal M}^2$
and $\hat{e}^9$ given in (\ref{m2mat}) and (\ref{elf}) respectively,
it turns out that $L^{\hat{r}}$
in (\ref{11super}) has non-vanishing component in $x^9$, especially,
$L^r_9 \neq 0$ except for $r=-$.  Thus, the super elfbein
does not satisfy the KK ansatz.

We would like to note that the obstacle for the dimensional reduction
depends on $h_- \theta^-$ (or $h_- {\mathcal P}_- \theta$)
which precisely corresponds to the components of eight broken
supersymmetries.  This is also the case in the construction of
the superspace for the 24 supersymmetric AdS$_4 \times CP^3$ background
through the dimensional reduction \cite{Gomis:2008jt}.  In some sense,
this kind of structural similarity may be expected naturally because
the IIA plane wave background is related to AdS$_4 \times CP^3$
via the Penrose limit.  We may guess that, if some fraction of the
supersymmetry was not broken along the direction of compactification,
the super elfbein would not lead to any problem in going down to
ten dimensions.

For the consistent KK reduction, the non-vanishing $L^r_9$ component of
the super elfbein should be eliminated.
As has been done also in the case of AdS$_4 \times CP^3$ background
\cite{Gomis:2008jt}, the way to eliminate it is to perform an appropriate
local Lorentz transformation in the plane tangential to the eleven
dimensional plane wave geometry.  Then, let us denote the transformed
super elfbein as $\widehat{E}^{\hat{r}}$ and consider the Lorentz
transformation,
\begin{equation}
\widehat{E}^{\hat{r}} = L^{\hat{s}} \Lambda_{\hat{s}}{}^{\hat{r}} \,.
\label{elftrans}
\end{equation}
The problem is to determine the transformation matrix
$\Lambda_{\hat{s}}{}^{\hat{r}}$ in such a way that
the component $\widehat{E}_9{}^r$ of $\widehat{E}^{\hat{r}}$ vanishes,
that is, $\widehat{E}_9{}^r = 0$.
It is not so difficult to solve this.
By the aid of the orthogonality condition,
\begin{align}
\Lambda_{\hat{r}}{}^{\hat{t}} \Lambda_{\hat{s}}{}^{\hat{u}}
\eta_{\hat{t}\hat{u}} = \eta_{\hat{r}\hat{s}} \,,
\label{ortho}
\end{align}
and the requirement of proper Lorentz transformation,
$\det \Lambda_{\hat{r}}{}^{\hat{s}} = +1$,
$\Lambda_{\hat{s}}{}^{\hat{r}}$ is uniquely determined as
\begin{gather}
\Lambda_9{}^9 = \frac{1}{\sqrt{1+v^2}} \,, \quad
\Lambda_r{}^9 =
  \frac{\eta_{rs} v^s}{\sqrt{1+v^2}} \,, \quad
\Lambda_9{}^r =
  -\frac{v^r}{\sqrt{1+v^2}} \,,
\notag \\
\Lambda_s{}^r = \delta_s^r
  -\frac{\sqrt{1+v^2}-1}{v^2 \sqrt{1+v^2}}
  \eta_{st} v^t v^r\,,
\label{vtrans}
\end{gather}
where we have defined
\begin{equation}
v^r \equiv \frac{L_9^r}{L_9^9} \,, \quad
v^2 \equiv \eta_{rs} v^r v^s \,.
\label{vr}
\end{equation}
Here $L_9^r=0$ and $L_9^9=1$ ($v^r=0$) when $h_- \theta^- = 0$.
Thus the
transformation matrix $\Lambda_{\hat{s}}{}^{\hat{r}}$ becomes the
unit matrix when the fermionic components corresponding to the
broken supersymmetry vanish.  One property of $v^r$ is that $v^-$ always
vanishes because $L^-_9=0$ as mentioned below of (\ref{10cov}), and thus
$v^2 = v^I v^I$.

The Lorentz transformation (\ref{elftrans}) with the transformation
matrix (\ref{vtrans}) is the one for the vector quantities.  In addition
to this, we should perform a corresponding Lorentz transformation
also for the spinor superfield $L$,
\begin{align}
\widehat{E}^{\hat{a}} = L^{\hat{b}} S_{\hat{b}}{}^{\hat{a}} \,.
\label{spintrans}
\end{align}
The transformation matrix $S$ is derived by making use of the relation
between the vector and the spinor representations of the Lorentz group,
\begin{align}
S^{T-1} \Gamma^{\hat{r}} S^T =
\Gamma^{\hat{s}} \Lambda_{\hat{s}}{}^{\hat{r}} \,.
\label{svrelation}
\end{align}
If we take the standard expression
$S^T = \exp \left( \frac{1}{4} \Gamma_{\hat{r}\hat{s}}
\epsilon^{\hat{r}\hat{s}} \right)$ with the transformation parameter
$\epsilon^{\hat{r}\hat{s}}$ and investigate the infinitesimal
transformation, it follows that $\Gamma_{r9}$ generates the Lorentz
transformation for the spinorial quantities.  Based on this, we can
obtain the explicit expression of $S$ for finite transformation as
\begin{align}
S & = \exp \left( {\frac{1}{2} \Gamma_{r9}^T \epsilon^{r9}} \right) \quad
{\rm with} \quad
\epsilon^{r9} = -\tan^{-1} |v| \frac{v^r}{|v|}
\notag \\
&= \frac{1}{\sqrt{2}(1+v^2)^{1/4}}
  \left( {\mathbf 1}_{32} \sqrt{ \sqrt{1+v^2} +1 }
       - \Gamma_{r9}^T \frac{v^r}{|v|} \sqrt{ \sqrt{1+v^2} -1 }
  \right) \,,
\label{strans}
\end{align}
where ${\mathbf 1}_{32}$ is the $32 \times 32$ unit matrix.

The transformed super elfbein, $\widehat{E}^{\hat{r}}$ and
$\widehat{E}^{\hat{a}}$, has the required form for the KK reduction.
Thus, we are now ready to get the ten dimensional superfields by
following the relation between the eleven and ten dimensional quantities
\cite{Duff:1987bx}.  First of all, the dilaton and dilatino superfield
are obtained as
\begin{align}
\Phi^{2/3} & = \widehat{E}_9^9
 = L^9_9 \sqrt{ 1+ v^2} \,,
\notag \\
\chi^a &= \Phi^{1/3} \widehat{E}_9^a
 = \Phi^{1/3} L_9^b S_b{}^a \,.
\label{10super1}
\end{align}
We note that, when the eleven dimensional spinorial quantity is
related to the ten dimensional one, a dilaton factor $e^{\phi/6}$ should
be multiplied for each spinor index such as
$\theta^a = e^{\phi/6} \theta^{\hat{a}}$.  This is basically due to the
necessity for having the canonical supersymmetry transformation rule
in ten dimensions.  However, the dilaton field is trivial in the IIA plane
wave background, and thus the dilaton factor does not appear in the above
expression.

As for the super zehnbein, we get
\begin{align}
E^r &= \Phi^{1/3} dZ^M \widehat{E}_M^r
\notag \\
  &= dZ^M ( \Phi^{1/3} L_M^s \Lambda_s{}^r - \Phi^{-1/3} L_M^9 L_9^r) \,,
\notag \\
E^a &= dZ^M (\Phi^{1/3} \widehat{E}^a_M
              - \Phi^{-1/3} \widehat{E}_M^9 \widehat{E}_9^a )
\notag \\
  &= dZ^M
     \left( \Phi^{1/3} L_M^b - \frac{\Phi^{-1/3}}{\sqrt{1+v^2}}
           (L^r_M v_r + L_M^9) L_9^b
     \right) S_b{}^a \,,
\label{10super2}
\end{align}
where $\Lambda_s{}^r$ and $S_b{}^a$ are the Lorentz transformation
matrices of (\ref{vtrans}) and (\ref{strans}).  We note that, although
it is clear from the form $dZ^M L^A_M$, the ten dimensional
super-covariant derivative (\ref{10cov}) is used in super elfbein
instead of the eleven dimensional one  (\ref{11-10cov}).  That is, we
use the super elfbein given in  (\ref{11super}) but with the replacement
of $\widehat{D} \theta$ by $D \theta$.  One may think that
$\hat{e}^9$ appearing in $L_M^9$ and the relation between the eleven and
ten dimensional super-covariant derivatives (\ref{11-10cov}) somehow
contributes to super zehnbein because we have $\hat{e}^9_- \neq 0$ from
(\ref{elf}) which corresponds to the non-vanishing R-R one-form gauge
field.  However, an explicit calculation shows that such part does not
contribute to the super zehnbein.  In fact, this should be the case
because the super zehnbein is neutral under the gauge transformation
associated with the R-R one-form gauge field.  Thus, it is understood
that, in the actual evaluation of (\ref{10super2}), $\hat{e}^9_\mu$ is
ignored and the super-covariant derivative is ten dimensional
one.

We turn to the three-form superfield $\widehat{B}$.  As one can see
from its expression (\ref{s3f}), it is a Lorentz scalar because it does
not contain any index in tangent space.  Therefore, the effect of
the local Lorentz transformation is just to replace the quantities in
its expression with the transformed ones, and what we get after the
transformation is
\begin{equation}
\widehat{B}
= \frac{1}{6} \hat{e}^{\hat{r}} \wedge \hat{e}^{\hat{s}} \wedge
    \hat{e}^{\hat{t}} \widehat{C}_{\hat{r}\hat{s}\hat{t}}
 + \int^1_0 d\lambda \; \bar{\theta}' \Gamma_{\hat{r}\hat{s}}
    \widehat{E}_\lambda \wedge \widehat{E}^{\hat{r}}_\lambda \wedge
    \widehat{E}^{\hat{s}}_\lambda \,,
\label{s3f2}
\end{equation}
where
the subscript $\lambda$ means the rescaling
$\theta \rightarrow \lambda \theta$
in the superfields as in (\ref{s3f})
and $\bar{\theta}'$ is the transformed fermionic coordinate given
by\footnote{We have used $SCS^T=C$, the property of the charge
conjugation matrix $C$ under the Lorentz transformation.}
\begin{align}
\bar{\theta}'=\bar{\theta} (S^T_\lambda)^{-1} \,,
\end{align}
with $S_\lambda^T = S^T |_{\theta \rightarrow \lambda \theta}$.
The first term on the right
hand side remains intact under the Lorentz transformation because it is
originated from the purely bosonic part of the closed four-form $H$,
$H_{\lambda=0}$, as can be seen
from (\ref{ht}), and the effect of Lorentz transformation with the
transformation matrix $\Lambda_{\hat{s}}{}^{\hat{r}}$ disappears at
$\lambda=0$.\footnote{Here, $\Lambda_{\hat{s}}{}^{\hat{r}}$ is also
understood
as the rescaled one through $\theta \rightarrow \lambda \theta$.}

Under the KK reduction, the three-form superfield
gives R-R three-form and NS-NS two-form gauge superfields in ten
dimensions.  Here, we restrict our attention to the NS-NS
superfield because it is relevant in constructing the superstring action.
If we denote it as $B= \frac{1}{2} dZ^M \wedge dZ^N B_{NM}$, then
$B_{NM}$ corresponds to $\widehat{B}_{9NM}$ \cite{Duff:1987bx} and
$B$ is given by
\begin{align}
B &= \frac{1}{2} dZ^M \wedge dZ^N \widehat{B}_{9NM}
\notag \\
 &=  \int^1_0 d\lambda
  \left(
        2 \bar{\theta}' \Gamma_{r9} \widehat{E}_\lambda \wedge
         \widehat{E}^r_\lambda
         \widehat{E}^9_{\lambda 9}
        +2 \bar{\theta}' \Gamma_{r9} \widehat{E}_{\lambda 9}
          \widehat{E}^r_\lambda
          \wedge \widehat{E}^9_\lambda
        + \bar{\theta}' \Gamma_{rs} \widehat{E}_{\lambda 9}
          \widehat{E}^r_\lambda
          \wedge \widehat{E}^s_\lambda
  \right)
\notag \\
 &=  \int^1_0 d\lambda
  \left(
        2 \bar{\theta}' \Gamma_{r9} E_\lambda \wedge E^r_\lambda
        + \Phi^{-1}_\lambda \bar{\theta}' \Gamma_{rs} \chi_\lambda
           E^r_\lambda \wedge E^s_\lambda
  \right) \,,
\label{b2}
\end{align}
where the three-form gauge field $\widehat{C}_{\hat{r}\hat{s}\hat{t}}$
does not contribute to $B$ because its field strength does not span
along $x^9$ direction in the plane wave background (\ref{pp-wave}).

\section{Type IIA superstring action on IIA plane wave background}
\label{stringaction}

We have obtained all the ten dimensional superfields necessary
for the construction of type IIA superstring action on the IIA
plane wave background, and are now ready to write down the action
containing all the 32 fermionic components.

The general form of type IIA Green-Schwarz superstring action is
\begin{align}
S_{\text{IIA}} = - \frac{1}{4\pi \alpha'} \int d^2 \sigma
 \sqrt{-h} h^{mn} \Pi^r_m \Pi^s_n \eta_{rs}
 + \frac{1}{2\pi \alpha'} \int B_2 \,,
\end{align}
where $\Pi^A_m$ and $B_2$ are the pullback of the super zehnbein and the
NS-NS two-form gauge superfield onto the string worldsheet given by
\begin{align}
\Pi^A_m &= \partial_m Z^M E^A_M \,,
\notag \\
B_2 &= \frac{1}{2} d^2 \sigma \epsilon^{mn}
   \partial_m Z^M \partial_n Z^N B_{NM} \,.
\end{align}
As for the worldsheet quantities,
$\sigma^m$ ($m=0,1$) is the worldsheet coordinate with the usual
notation
\begin{align}
\sigma^0 = \tau \,, \quad \sigma^1 = \sigma \,,
\end{align}
$h^{mn}$ is the worldsheet metric, and the anti-symmetric
tensor $\epsilon^{mn}$ follows the convention $\epsilon^{01}=+1$.

If we now plug the expression of (\ref{b2}) for the NS-NS two-form
superfield into the superstring action, then we have
\begin{align}
S_{\text{IIA}} =& -\frac{1}{4\pi \alpha'} \int d^2 \sigma
 \sqrt{-h} h^{mn} \partial_m Z^M E^r_M \partial_n Z^N E^s_N  \eta_{rs}
\notag \\
&  +\frac{1}{2\pi \alpha'} \int d^2 \sigma
    \int^1_0 d\lambda
      \left(
        2 \bar{\theta}' \Gamma_{r9} E_\lambda \wedge E^r_\lambda
        + \Phi^{-1}_\lambda \bar{\theta}' \Gamma_{rs} \chi_\lambda
           E^r_\lambda \wedge E^s_\lambda
      \right) \,,
\end{align}
where the expressions for various superfields are given in
(\ref{10super1}) and (\ref{10super2}).  Thus, we have achieved our
goal of constructing the complete type IIA Green-Schwarz superstring
action on the IIA plane wave background. However, the action depends
implicitly on other expressions such as the super elfbein of
(\ref{11super}) and the Lorentz transformation matrices in
(\ref{vtrans}) and (\ref{strans}).   To facilitate the better
understanding of the derived results, we perform a little bit more
manipulation for the superstring action and give a summary of
related expressions.

The superstring action is composed of the kinetic and the Wess-Zumino
term:
\begin{align}
S_{\text{IIA}} = S_{\text{kin}} + S_{\text{WZ}} \,.
\label{iiaaction}
\end{align}
Let us first consider the kinetic term.  Then (\ref{10super1}) and
(\ref{10super2}) allow us to express it as follows.
\begin{align}
S_{\text{kin}} =& -\frac{1}{4\pi \alpha'} \int d^2 \sigma
 \sqrt{-h} h^{mn} \partial_m Z^M \partial_n Z^N (-1)^{\langle M,N \rangle}
 ( \Phi^{2/3} L_M^r L_N^s \eta_{rs}
\notag \\
 &-\Phi^{-2/3} L_M^r L_N^s L_9^t L_9^u \eta_{rt} \eta_{su}
  - 2 \Phi^{-2/3} L_M^r L_N^9 L_9^s L_9^9 \eta_{rs}
   +\Phi^{-2/3} L_M^9 L_N^9 L_9^r L_9^s \eta_{rs}) \,,
\label{kin}
\end{align}
where the expressions of (\ref{vtrans}) for the Lorentz transformation,
and the properties of the Lorentz transformation matrices,
(\ref{ortho}) and (\ref{svrelation}), have been used.  The symbol
$\langle M,N \rangle$ means that $\langle M,N \rangle =1$ when
both of $M$ and $N$ are spinorial and $\langle M,N \rangle =0$
otherwise.  Similar manipulation for the Wess-Zumino term leads us
to have
\begin{align}
S_{\text{WZ}}
=& \frac{1}{2\pi \alpha'} \int d^2 \sigma \int^1_0 d\lambda \;
 ( \; \bar{\theta} \Gamma_{rs} L_{\lambda 9} L^r_\lambda
           \wedge L^s_\lambda
  + 2 \bar{\theta} \Gamma_{r9} L_\lambda \wedge L^r_\lambda
            L^9_{\lambda 9}
  + 2 \bar{\theta} \Gamma_{rs} L_\lambda \wedge L^r_\lambda
           L^s_{\lambda 9}
\notag \\
& + 2 \bar{\theta} \Gamma_{r9} L_{\lambda 9} L^r_\lambda \wedge
          L^9_\lambda
  - 2 \bar{\theta} \Gamma_{r9} L_\lambda \wedge L^9_\lambda
           L^r_{\lambda 9} \; ) \,,
\label{wzterm}
\end{align}
where the wedge product is understood as the pullback version, that is,
for example
\begin{align}
L^r \wedge L^s \equiv \epsilon^{mn} \partial_m Z^M L_M^r
\partial_n Z^N L_N^s \,.
\end{align}
Explicit expressions for the various quantities appearing in
the action are given by
\begin{align}
\Phi^{2/3} &= \sqrt{ (L_9^9)^2 + L_9^r L_9^s \eta_{rs}} \,,
\notag \\
dZ^M L_M^r &= e^r
 -2  \sum^{15}_{n=0} \frac{1}{(2n+2)!} \bar{\theta}\Gamma^r
   {\mathcal M}^{2n} D \theta \,,
\notag \\
dZ^M L_M^9 &= -2  \sum^{15}_{n=0} \frac{1}{(2n+2)!}
    \bar{\theta}\Gamma^9 {\mathcal M}^{2n} D \theta \,,
\notag \\
dZ^M L_M &= \sum^{16}_{n=0} \frac{1}{(2n+1)!} {\mathcal M}^{2n}
    D \theta \,,
\notag \\
L_9^r &= - \frac{\mu}{3} \sum^{15}_{n=0} \frac{1}{(2n+2)!}
    \bar{\theta}\Gamma^r {\mathcal M}^{2n} \Gamma^{-4} h_- \theta^- \,,
\notag \\
L_9^9 &= 1 - \frac{\mu}{3} \sum^{15}_{n=0} \frac{1}{(2n+2)!}
    \bar{\theta}\Gamma^9 {\mathcal M}^{2n} \Gamma^{-4} h_- \theta^- \,,
\notag \\
L_9 &= \frac{\mu}{6} \sum^{16}_{n=0} \frac{1}{(2n+1)!} {\mathcal M}^{2n}
    \Gamma^{-4} h_- \theta^- \,,
\label{fields}
\end{align}
where the ten dimensional super-covariant one form $D \theta$ and the
$32 \times 32$ matrix ${\mathcal M}^2$ are given in (\ref{10cov}) and
(\ref{m2mat}) respectively.

\subsection{Light-cone gauge fixed action}

The superstring action (\ref{iiaaction}) is a complete action containing
all the 32 fermionic coordinates.  We now take the fermionic
and the bosonic light-cone gauge choices and obtain the superstring
action in the light-cone gauge.

We first fix the fermionic $\kappa$-symmetry by taking the usual
$\kappa$-symmetry light cone gauge
\begin{align}
\Gamma^- \theta=0 \quad (\theta^- = {\mathcal P}_- \theta = 0 ) \,.
\end{align}
Under this choice, it follows immediately that
\begin{align}
L^r_9=0 \,, \quad L^9_9=1 \,, \quad
L_9=0 \,, \quad \Phi=1 \,,
\label{lcs1}
\end{align}
as can be seen from (\ref{fields}).
The super-covariant one form $D\theta$ of (\ref{10cov}) is simplified
as
\begin{align}
D\theta = d \theta
  - \frac{\mu}{4} e^-
      \left(\Gamma^{123}+\frac{1}{3} \Gamma^{49} \right)\theta  \,.
\label{lccov}
\end{align}
As for the matrix ${\mathcal M}^2$ of (\ref{m2mat}), it simply vanishes
due to the presence of projection operator ${\mathcal P}_-$ in every
terms.  This fact leads to a pretty much simplification for the remaining
superfields of (\ref{fields}) as follows.
\begin{align}
dZ^M L_M^r = e^r - \bar{\theta}\Gamma^r D \theta \,, \quad
dZ^M L_M^9 = - \bar{\theta}\Gamma^9 D \theta \,, \quad
dZ^M L_M =  D \theta \,,
\label{lcs2}
\end{align}
with $D\theta$ of (\ref{lccov}).

If we plug the above expressions from (\ref{lcs1}) to (\ref{lcs2})
into the superstring action (\ref{iiaaction}) and use the bosonic
zehnbein of (\ref{zehn}), then we obtain the $\kappa$-symmetry
fixed superstring action as
\begin{align}
S_{\text{IIA}}
=& -\frac{1}{4 \pi \alpha'} \int d^2 \sigma \sqrt{-h} h^{mn}
  \bigg[ 2 \partial_m X^+ \partial_n X^-
    + \partial_m X^I \partial_n X^I
    - A(X^I) \partial_m X^- \partial_n X^-
                         \nonumber \\
   & + 2 \partial_m X^- \bar{\theta} \Gamma^+ \partial_n \theta
      + \frac{\mu}{2} \partial_m X^- \partial_n X^-
       \bar{\theta} \Gamma^+
         \left( \Gamma^{123} + \frac{1}{3} \Gamma^{49}
         \right) \theta
  \bigg]
                         \nonumber \\
  &  - \frac{1}{2 \pi \alpha'} \int d^2 \sigma
   \epsilon^{mn} \partial_m X^- \bar{\theta} \Gamma^{+9}
    \partial_n \theta ~.
\label{kfix-action}
\end{align}

We now turn to the bosonic light-cone gauge.  The equation of motion
for $X^-$ is harmonic, and thus the light-cone gauge, $X^- \propto
\tau$, is allowed.  Let us take the following light-cone gauge
choice
\begin{align}
X^- = \alpha' p^- \tau \,,
\end{align}
where $p^-$ is the total momentum conjugate to $X^+$.  With this choice,
the worldsheet diffeomorphism can be consistently fixed as $\sqrt{-h}=1$,
$h_{\sigma\tau}=0$, which allow us to fix other worldsheet metric
components as $h_{\tau\tau}=-1$ and $h_{\sigma\sigma}=1$.
Then the $\kappa$-symmetry fixed superstring action (\ref{kfix-action})
is further simplified, and the superstring action in the light-cone
gauge, $S_{\text{LC}}$, is obtained finally as
\begin{align}
S_{\text{LC}}
 =& - \frac{1}{4 \pi \alpha'} \int  d^2 \sigma
 \Bigg[ \eta^{mn} \partial_m X^I \partial_n X^I
      + \frac{m^2}{9} (X^i)^2
      + \frac{m^2}{36} (X^{i'})^2
                       \nonumber \\
 & - \bar{\theta} \Gamma^+ \partial_\tau \theta
     + \bar{\theta} \Gamma^{+9} \partial_\sigma \theta
     - \frac{m}{4} \bar{\theta} \Gamma^+
        \left( \Gamma^{123} + \frac{1}{3} \Gamma^{49} \right)
        \theta
  \Bigg] ~,
\label{lcaction}
\end{align}
where the fermionic coordinate has been rescaled as $\theta
\rightarrow \theta / \sqrt{2 \alpha' p^-}$ and $m$ defined as $m
\equiv \mu \alpha' p^-$ is a mass parameter for the worldsheet
variables.  This is the action considered previously
\cite{Sugiyama:2002tf,Hyun:2002wu}, and thus shows that our complete
superstring action of (\ref{iiaaction}) satisfies a basic
consistency check.

\section{Conclusion}
\label{concl}

We have constructed the complete type IIA Green-Schwarz superstring
action on the  ten dimensional IIA plane wave background with 24
supersymmetries. As a consistency check, we have obtained the
superstring action in the light-cone gauge and shown that it is
exactly the same as that considered previously.

Having the complete action containing all the 32 fermionic
components, we can study the various possible superstring
configurations by taking an appropriate $\kappa$-symmetry fixing
condition.  Especially interesting thing is the configuration whose
correct quantum description requires the fermionic components
corresponding to the broken supersymmetries.  As noted in the
introduction, in the case of type IIA superstring on AdS$_4 \times
CP^3$, such fermionic components are crucial for the description of
superstring moving only in AdS$_4$ space \cite{Gomis:2008jt}.  Under
the Penrose limit relating the AdS$_4 \times CP^3$ space to the type
IIA plane wave background, the string configuration embedded only in
AdS$_4$ space would correspond to that spanned in the space
parametrized by $x^i$.  Since the superstring on the type IIA plane
wave background is rather simpler than that on AdS$_4 \times CP^3$
space, we expect that we can understand such superstring
configuration more clearly.

Another interesting issue that may be considered with the complete
superstring action is the realization of worldsheet supersymmetry.  As
has been shown in \cite{Hyun:2002wu}, the superstring action in the
light-cone gauge (\ref{lcaction}) has ${\mathcal N}=(4,4)$ worldsheet
supersymmetry.  Among the 16 fermionic components eliminated by the
$\kappa$-symmetry light cone gauge $\Gamma^- \theta=0$, 8 components
correspond to the broken supersymmetries.  Other 8 components are
responsible for the worldsheet supersymmetry which basically stems
from the fact that the supersymmetry transformation parameter
$h_+ \epsilon^-$ discussed at the end of Sec.~\ref{10dbosgeo} does
not satisfy the $\kappa$-symmetry light-cone gauge.  At this point,
one may be curious about the worldsheet supersymmetry realized after
taking another consistent $\kappa$-symmetry fixing condition.  In
the case of maximally supersymmetric backgrounds, there would be nothing
special and we would get unique structure on worldsheet supersymmetry.
However, it seems that there would be some change in the supersymmetry
structure in less supersymmetric backgrounds such as the present IIA
plane wave background.  The work on this issue is in progress, and will
be reported elsewhere.

\section*{Acknowledgments}

HS would like to thank Makoto Sakaguchi for helpful discussion and
informing his work relevant to this work. This work was supported by
the National Research Foundation of Korea (NRF) grant funded by the
Korea government (MEST)  with the Grants No.~2009-0085995 (JP),
2008-331-C00071 (HS) and 2005-0049409 (JP) through the Center for
Quantum Spacetime (CQUeST) of Sogang University. JP is also
suppported by POSTECH BSRI fund with the Grants No. 4.0007999.01 and
appreciates APCTP for its stimulating environment for research.


\appendix
\section{Notation}
\label{app1}

The eleven dimensional quantities and indices are denoted basically
by using hat to distinguish from those in ten dimensions.

In eleven dimensions, $\hat{M}$, $\hat{N}$, \dots
($\hat{A}$, $\hat{B}$, \dots) are the
target (tangent) superspace indices.  Each superspace
index is the composition of two types of indices such as
$\hat{M} = (\hat{\mu},\hat{\alpha})$
($\hat{A} = (\hat{r},\hat{a})$).  Convention for each index and
related indices is as follows:
\begin{align*}
&\hat{r},\hat{s},\dots = +,-,1,\dots,9 &&
   \text{11D tangent space-time vector indices}
\\
&\hat{\mu},\hat{\nu},\dots = +,-,1,\dots,9 &&
   \text{11D target space-time vector indices}
\\
&\hat{a},\hat{b},\dots = 1,\dots,32 &&
   \text{11D tangent space-time spinor indices}
\\
&\hat{\alpha},\hat{\beta},\dots = 1,\dots,32 &&
   \text{11D target space-time spinor indices}
\\
&\hat{I},\hat{J},\dots = 1,\dots,9 &&
   \text{$SO(9)$ vector indices} \quad (\hat{I}=(\hat{i},\hat{i}'))
\\
&\hat{i},\hat{j},\dots = 1,2,3 &&
\\
&\hat{i}',\hat{j}',\dots = 4,\dots,9 &&
\end{align*}
The metric $\eta_{\hat{r}\hat{s}}$ for the tangent space-time follows the most plus convention. Since the light-cone coordinate is defined as
\begin{align}
x^\pm \equiv \frac{1}{\sqrt{2}} ( x^{11} \pm x^0 )\,,
\end{align}
$\eta_{+-}=1$ with the spatial part
$\eta_{\hat{I}\hat{J}}=\delta_{\hat{I}\hat{J}}$.

In ten dimensions, $M$, $N$, \dots ($A$, $B$, \dots) are the
target (tangent) superspace indices.  Similar to the eleven dimensional
case, $M = (\mu,\alpha)$ ($A = (r,a)$) with the following convention.
\begin{align*}
&r,s,\dots = +,-,1,\dots,8 &&
   \text{10D tangent space-time vector indices}
\\
&\mu,\nu,\dots = +,-,1,\dots,8 &&
   \text{10D target space-time vector indices}
\\
&a,b,\dots = 1,\dots,32 &&
   \text{10D tangent space-time spinor indices}
\\
&\hat{\alpha},\hat{\beta},\dots = 1,\dots,32 &&
   \text{10D target space-time spinor indices}
\\
&I,J,\dots = 1,\dots,8 &&
   \text{$SO(8)$ vector indices} \quad (I=(i,i'))
\\
&i,j,\dots = 1,2,3,4 &&
\\
&i',j',\dots = 5,6,7,8 &&
\end{align*}

\end{document}